# An Open-Source Multi-functional Testing Platform for Optical Phase Change Materials

*Cosmin-Constantin Popescu, Khoi Phuong Dao, Luigi Ranno, Brian Mills[†], Louis Martin, Yifei Zhang, David Bono. Brian Neltner, Tian Gu, Juejun Hu\**

Department of Materials Science & Engineering, Massachusetts Institute of Technology Cambridge, MA 02139, USA

E-mail: hujuejun@mit.edu

[†] Draper Scholar, The Charles Stark Draper Laboratory, 555 Technology Square, Cambridge, MA 02139, USA

*Kiumars Aryana, William M. Humphreys, Hyun Jung Kim\**

NASA Langley Research Center, Hampton, VA 23681, USA

E-mail: hyunjung.kim@nasa.gov

*Steven Vitale, Paul Miller, Christopher Roberts*

Advanced Materials and Microsystems Group, MIT Lincoln Laboratory Lexington, MA 02421, USA

*Sarah Geiger, Dennis Callahan, Michael Moebius*

The Charles Stark Draper Laboratory, Inc., Cambridge, MA 02139, USA

*Myungkoo Kang, Kathleen Richardson*

CREOL, The College of Optics & Photonics University of Central Florida Orlando, FL 32816, USA

*Carlos A. Ríos Ocampo*

Department of Materials Science & Engineering University of Maryland College Park, MD 20724, USA

Owing to their unique tunable optical properties, chalcogenide phase change materials are increasingly being investigated for optics and photonics applications. However, *in situ* characterization of their phase transition characteristics is a capability that remains inaccessible to many researchers. In this article, we introduce a multi-functional silicon microheater platform capable of *in situ* measurement of structural, kinetic, optical, and thermal properties of these materials. The platform can be fabricated leveraging industry-standard silicon foundry manufacturing processes. We fully open-sourced this platform, including complete hardware design and associated software codes.

## 1. Introduction

Phase change materials (PCMs) are a class of chalcogenides featuring a large change in electrical and optical properties between their amorphous and crystalline phases. For decades, these materials have been used in optical data storage as well as electronic memories with exemplary compositions such as $Ge_2Sb_2Te_5$ (GST) and $Ag_xIn_ySb_2Te$ (AIST)[1,2]. In the past decade, they have been gaining increasing attention due to their potential uses in photonics[3–9]. The key trait that underpins this emerging field is the nonvolatile, large optical property contrast between their amorphous and crystalline states. This trait foresees a cohort of energy-efficient, ultra-compact reconfigurable photonic devices for applications such as neuromorphic optical computing[10–14], photonic memories[15–18], active metasurfaces[19–25], tunable filters[26–28], reflective displays[29–32], thermal camouflage[33–35], programmable photonic circuits[36–39], and more[40–42].

Despite the surge of interest in their optical applications, characterization of PCMs in environments relevant to photonic device settings, quite distinct from electronic memory configurations, is scarce. In photonic devices, electrothermal switching using resistive microheaters, which avoids non-uniform crystallization due to filamentation[43], is the preferred approach for controlling the phase structure in PCMs. Although a number of microheater platforms made of metals[44,45], transparent conducting oxides[46,47], doped Si[48–50], and graphene[51–53] have been demonstrated for this purpose, the phase transition characteristics of PCMs and their corresponding structural and thermal properties have not been systematically investigated. In addition, the failure mechanisms of PCMs in the context of photonic applications, which dictate their reliability and cycling endurance, have not been studied[54]. Moreover, photonic applications also stipulate performance requirements distinct from those in electronic memories, such as low optical attenuation and, in some cases, moderate crystallization speed[43]. Therefore, a series of new PCM compositions tailored for photonic applications exemplified by $Ge_2Sb_2Se_4Te$ (GSST)[55–57], $Sb_2S_3$[58], $Sb_2Se_3$[59], $Ge_2Sb_2Te_3S_2$[60], and $In_3SbTe_2$[61,62] have been developed in recent years. Compared to the archetypal GST family, the structural, kinetic, thermal, and cycling characteristics of these emerging optical PCMs are much less understood. A platform that enables comprehensive characterization of PCMs in environments pertinent to photonic device applications is thus desired.

In this article, we detail the design and implementation of a multi-functional platform for optical PCM characterization. The platform uses doped silicon-on-insulator (SOI) microheaters to actuate phase transition in PCMs. We choose SOI heaters because they form the backbone of most photonic integrated circuits, their fabrication is broadly accessible to the community through commercial photonic foundries, their infrared transparency allows transmissive optical interrogation, and finally they can further act as a probe for local temperature measurement via Raman thermometry. As we will show later, the platform integrates multiple *in situ* characterization methods to reveal a rich wealth of information regarding structural, kinetic, optical and thermal properties of PCMs. Finally, we foresee that the platform, which can be readily coupled with combinatorial deposition[63,64], can also facilitate high-throughput screening of PCMs to expedite new PCM discovery.

## 2. Testing platform setup

The SOI microheaters are fabricated at the MIT Lincoln Laboratory Microelectronics Laboratory. We note that similar fabrication processes are also available through standard multi-project wafer shuttle runs in most photonic foundries. A schematic of the fabrication sequence is illustrated in **Figure 1**a. Wafers with a 220 nm SOI layer and 3 µm buried oxide were used as the starting substrates. The microheaters were n-doped with implanted phosphorous ions with 80 kV and a dose of $10^{16}$ cm$^{-2}$ followed by rapid thermal annealing for 10 s at 1000 °C (step 2 in Figure 1a). 10 nm of $SiO_2$ was grown on the device and an etch to the doped Si regions was performed. An adhesion layer of 10 nm Ti and 20 nm TiN was deposited on the exposed Si, followed by an etch back to define the contact regions (step 3). 350 nm of Al were deposited as the contacts (step 4). When needed, the chips were backside polished to facilitate transmissive optical measurements. Prior to PCM deposition, the chip was patterned with photolithography using 2 µm AZ nLOF 2020 resist and developed with AZ 300 MIF developer. The PCM was then deposited via thermal evaporation following established protocols[56], followed by pattern lift-off in acetone (step 5). The PCM films can be lithographically patterned to introduce optical functions, or to mitigate morphology-dependent failure mechanisms (e.g., liquid-phase morphological instability or dewetting). The PCM was subsequently capped via atomic layer deposition in 20 nm $Al_2O_3$ (at 110 °C in a Unitronics ALD system) and then further encapsulated in a thick reactive sputtered $SiN_x$ (deposited using Si targets with gas flow rates $N_2$:Ar 6:6 sccm at 3 mTorr pressure on an AJA Orion 5 system) protective layer (steps 6 and 7). The deposited encapsulation layers were removed from above the metal contact pads using reactive ion etching (RIE), specifically $SF_6$ and Ar plasma with a hard-baked AZ 3312 resist mask. After the etch-back, the remaining resist was removed via $O_2$ plasma. The microheaters are connected to a printed circuit board via wire bonding using an MEI model 1204 D ball bonder (step 8) to complete the microheater fabrication and packaging processes. It is preferable but not necessary that the PCB contact pads used for wire bonding are of the same metal as the connecting wire (i.e. Au-Au or Al-Al) for easier bonding.

**Figure 2** shows a schematic of the experimental setup. In particular, Figure 2b shows the electronic circuit for pulse amplification, where the microheater is denoted as R1. An IRF 510 power MOSFET transistor (M1 in Figure 2b) is used as a switch, and two bypass capacitors (C1 and C2 in Figure 2b) (a ceramic 1 µF capacitor and an electrolytic 470 µF capacitor for 50 V, respectively) are connected from the power supply to ground to remove high frequency noise. A 1N4745A Zener diode (D1) and 100 Ω resistor (R2) are connected as shown in the diagram at the gate of the transistor. The diode protects against voltages above 16 V from being applied to the gate while the series resistor R2 limits voltage gain at high oscillation frequencies. These are connected and encased in a Pomona Electronics 3234 shielded box with 4 BNC connectors. A power supply Keithely 2200-60-2 is used as DC voltage source (V1) and a function generator Agilent 33250 A (V2) is used to apply voltage pulses at the transistor gate with a rise/fall time of 5 ns. The communication to the power supply was done via USB port and the communication to the function generator was done via USB to RS 232 with null modem (for more details check the handshake selection for RS232 in the manual). Both instruments received commands via SCPI code which can be found in their respective manuals[65,66]. The waveform data collection was performed using a DPO 2014 Tektronix oscilloscope. The system is controlled via a MATLAB script (available through Github[67]) which allows synchronization with other instruments. During continuous cycling tests, a delay time between pulses (up to 30 s) is applied to allow for complete heat removal from the regions surrounding the microheaters. This prevents over-heating of the devices and damage to the metal contacts.

Figure 2c-e shows exemplary measured voltage waveforms at three different points in the circuit (labeled in Figure 2b). Ringing can be observed at ramp-up and ramp-down of the voltage pulses, although its impact on the heater temperature transient is negligible provided that the ringing oscillation period is considerably smaller than the thermal time constant of the heater. It is noted that the function generator is specified for a 50 Ω load at the output port following a 50 Ω source impedance. In our case, the function generator connects to a transistor whose gate to source resistance is much higher. As a result, the transistor gate acts as a circuit break, leading to doubling of the voltage at the gate. Therefore, in the examples shown in Figure 2c, the functional generator is nominally configured to output 4 V voltage pulses, which however results in an 8 V bias output at the transistor gate.

The thermal response of SOI heater platform to voltage pulses was modeled via the finite element method (FEM) using the COMSOL Multiphysics package. In the COMSOL model, experimentally assessed temperature-dependent doped Si conductivity (Figure S1) and PCM thermal properties[68] were used. Other material parameters were quoted from the COMSOL Multiphysics® v6.1 database[69]. In the simulations, a transient model was used to model amorphization while a steady-state model was employed for crystallization, since the crystallization pulse is much longer than the thermal time constant of the heater. For amorphization, a top-hat 32 V amplitude and 12 µs duration voltage pulse was applied across the heater. For steady state simulations, a constant 17 V bias was applied. The system started at 298 K and a constant temperature boundary condition was enforced to the bottom Si substrate. A convection heat flux was applied to the top surface using a heat transfer coefficient of 20 W m$^{-2}$ K$^{-1}$ [70], although our results indicate that the convective heat removal is negligible in comparison to conductive heat dissipation through the substrate.

The thermal simulation results are plotted in Figure 1d, showing the temperature as a function of time during amorphization, and in Figure 1e mapping the steady-state temperature profile across the heater during crystallization. Simulation results for 200 and 100 µm sized square heaters on 3 µm of buried oxide (BOX) along with result for a 100 µm sized heater on 1 µm BOX are provided. The voltages applied were chosen so that the maximum temperatures at the center of the heaters are nearly identical. The 200 and 100 µm sized heaters with 3 µm BOX exhibit similar temperature ramp and decay characteristics, whereas the heater on 1 µm BOX shows a much faster temperature transient response. The observation implies that heat removal occurs mostly through the substrate, and in-plane heat dissipation is negligible. The fitted thermal decay time constant for the 100 µm and 200 µm heater on 3 µm BOX is 7.2 µs and 7.1 µs, respectively, and 1.9 µs for the 100 µm heater on 1 µm of oxide. Characterization of PCMs with rapid crystallization kinetics can thus be performed using SOI wafers with thin BOX.

The PCM-integrated microheater arrays can be monitored *in situ* using non-contact optical techniques such as optical microscopy, micro-Fourier transform infrared (micro-FTIR) spectroscopy and micro-Raman spectroscopy. These techniques can be used to extract an extensive set of information including PCM morphology, phase composition, local temperature, optical constants, phase transition kinetics, and endurance, as we will discuss later. The devices can also be analyzed *ex situ* using destructive methods such as transmission electron microscopy (TEM) to reveal spatially resolved PCM composition and structural information, which proved a powerful tool to understand failure mechanisms of PCMs (detailed analysis to be published in a separate paper). For optical imaging, an AmScope AF205 autofocus camera was used for image acquisition. For white balance, the aluminum pads were used with resulting sensitivity values of R 15, G 32, B 68. A Thermo Scientific™ Nicolet™ iS50 spectrometer with a microscope using a Reflachromat™ condenser and a HgCdTe detector was used for transmission micro-FTIR. An inVia Renishaw spectrometer using a 50X objective (NA = 0.5), optical resolution 0.96 μm, was used for Raman spectroscopy measurements. The laser for Raman spectroscopy operates at 785 nm. In Raman measurements, it is important to ensure that the laser excitation power is sufficiently small to prevent optically induced structural changes.

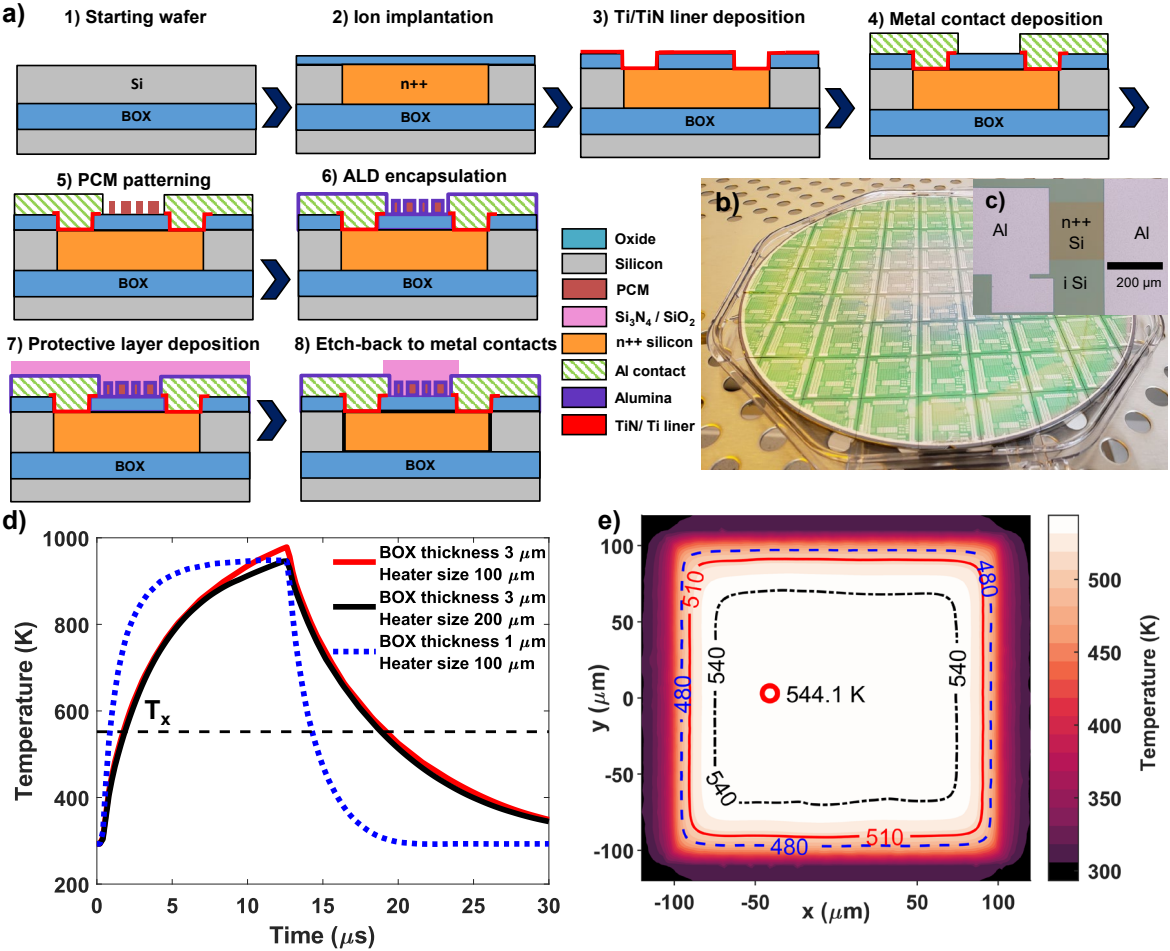

**Figure 1** a) Fabrication sequence for the PCM-integrated SOI heaters. b) A photo of a 200 mm wafer with the heater arrays. c) Top-view optical micrograph of a microheater. d) Simulated transient temperature responses at the center of the heater for 3 and 1 μm buried oxide for 12 μs amorphization pulses of 17 V (solid red), 32 V (solid black) and 24 V (dashed blue). e) Transverse, steady-state, temperature distribution across a 200 μm microheater resulting from a DC voltage, being applicable for crystallization. The red circle highlights the point of maximum temperature, with its position being influenced by the meshing of the heater in an area with low local temperature variation.

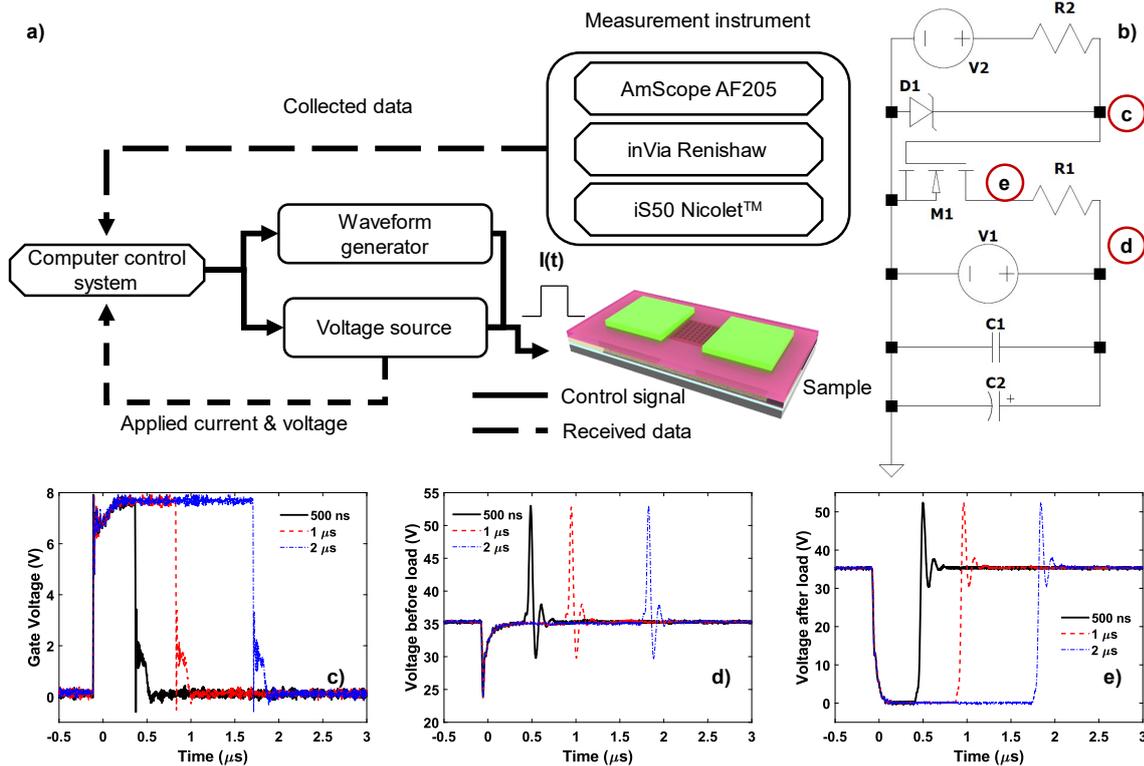

**Figure 2** a) Block diagram of the multi-functional characterization platform. B) Electric circuit diagram for high voltage pulse generation. c) Voltage measurements from an oscilloscope across the load measured c) at the transistor gate; d) before the load; and e) after the load.

## 3. Multi-modal PCM characterization

Here we take a GSST-integrated heater array to illustrate the repertoire of *in situ* analysis capabilities empowered by the platform. We first use transmissive FTIR to characterize the optical properties of PCM. Figure 3a presents the FTIR spectra of the GSST-integrated heater over multiple cycles. The spectra can be conveniently used as input to multilayer transmission models to fit the optical constants of PCMs at various switching states.

Phase composition of the PCM is then quantified via Raman spectroscopy. Figure 3b shows the Raman spectra collected on the device. The peak at 120 cm$^{-1}$ is primarily linked to vibrations of the Ge-6Se octahedra characteristic of ordered crystalline phases, and the crystalline phase fraction can be calculated by fitting the Raman spectra with Raman active vibrational modes of GSST[71]. From Figure 3b we also see that GSST in the as-fabricated device is partially crystallized, evidenced by the appearance of the Ge-6Se octahedra peak and likely caused by several heat treatment steps involved in the fabrication process.

In addition to determining the phase structures of PCM, Raman spectroscopy also enables non-contact, self-calibrating, and in-line probing of local temperatures at the SOI heater surface with micron-scale lateral resolution and ± 10 °C accuracy. This is accomplished by tracking the temperature-dependent spectral shift of the single crystal Si Raman peak[72]. In our experiment, we observed a clear red shift and subsequent splitting of the Si Raman peak into two as the applied DC voltage increases (Figure 3c inset). At low voltages, Raman peaks from the SOI heater and the Si substrate coincide due to a negligible heating effect. At higher voltages, the redshift of the SOI Raman peak is used to extract the heater temperatures plotted in Figure 3c, which also agree well with our FEM simulations. We note that while here we use standard Raman microscopy to measure steady-state temperatures at the heater, it is also possible to leverage ultrafast Raman thermometry to quantify the transient temperature evolution at sub-nanosecond time scale[73].

Combining the phase composition and temperature information enables quantitative evaluation of the time-temperature-transformation (TTT) diagram in PCMs, which is of vital importance not only to the understanding of phase transformation kinetics but also to PCM switching process optimization[71]. Specifically, high-throughput isothermal PCM crystallization experiments can be performed on the microheater array. The crystallization temperatures are calibrated using Raman *during* crystallization and the phase composition assessed via Raman fitting *after* crystallization. As an example, Figure 3d shows a measured TTT diagram of GSST. In the diagram, each point corresponds to one isothermal crystallization experiment on a microheater for a given heat treatment temperature and duration. The diagram points to a nose temperature of ~ 820 K, implying that crystallization of GSST should ideally be performed at this temperature to expedite the transition. The TTT diagram can be further refined by adding more data points to yield more accurate kinetics, an advantage afforded by the high-throughput testing capability on our platform.

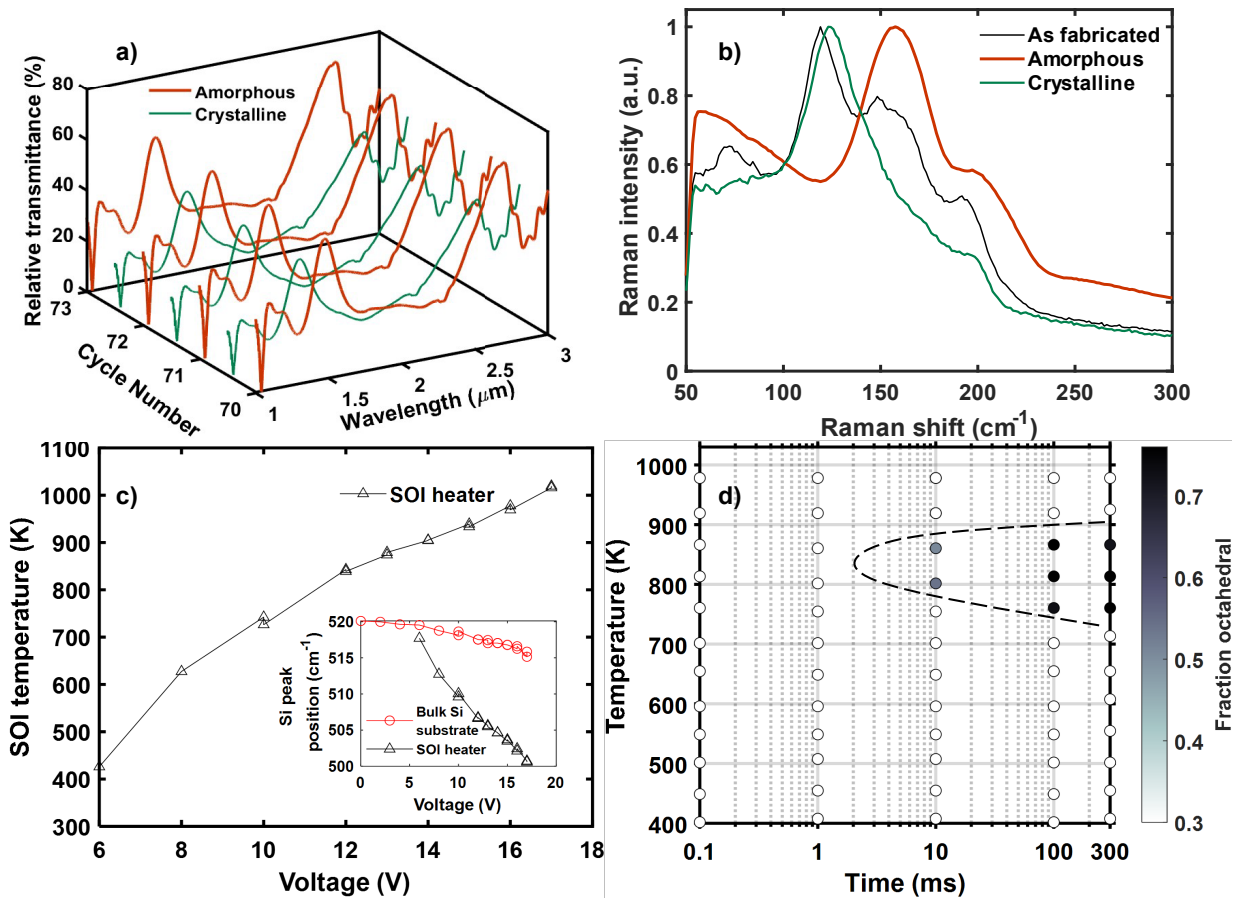

**Figure 3** a) Transmittance of a PCM-integrated SOI heater measured *in situ* via FTIR. b) Raman spectra of the PCM (GSST) from an as-fabricated device (black curve), after re-crystallization (green curve), and after re-amorphization (red curve). c) SOI heater temperature measured using Raman spectroscopy as a function of applied voltage at steady state. Inset shows the Si Raman peak positions for the SOI heater and bulk Si substrate vs. applied voltage across the heater. d) Measured TTT diagram of GSST (data reproduced from Vitale *et al.*[71]). The dashed line encircles the crystallization region.

The cycling behavior of the PCM is investigated through optical microscopy. We have set up an automated collection system and an image processing algorithm (available through Github [67]). The system records micrographs of the sample across three color channels (Red R,

Green G, and Blue B) of the camera, performs background subtraction (filtering out the non-PCM covered areas by setting a contrast threshold), normalizes the brightness level, and outputs the optical contrast evolution over cycles. The optical contrast is characterized by the differential mean analysis (DMA) parameter $\Delta \text{DMA} = \left[\frac{I_{PCM}}{I_{Bckg}}\right]_{Cr} - \left[\frac{I_{PCM}}{I_{Bckg}}\right]_{Am}$. Here $I$ represents the optical intensity in one of the color channels, the subscripts *PCM* and *Bckg* denote the intensity from the PCM film and the background (non-PCM covered region) respectively, and the subscripts *Cr* and *Am* refer to quantities associated with crystalline and amorphous states, respectively. The evolution of DMA over cycles provides a direct measure of the endurance as well as consistency of the PCM cycling process.

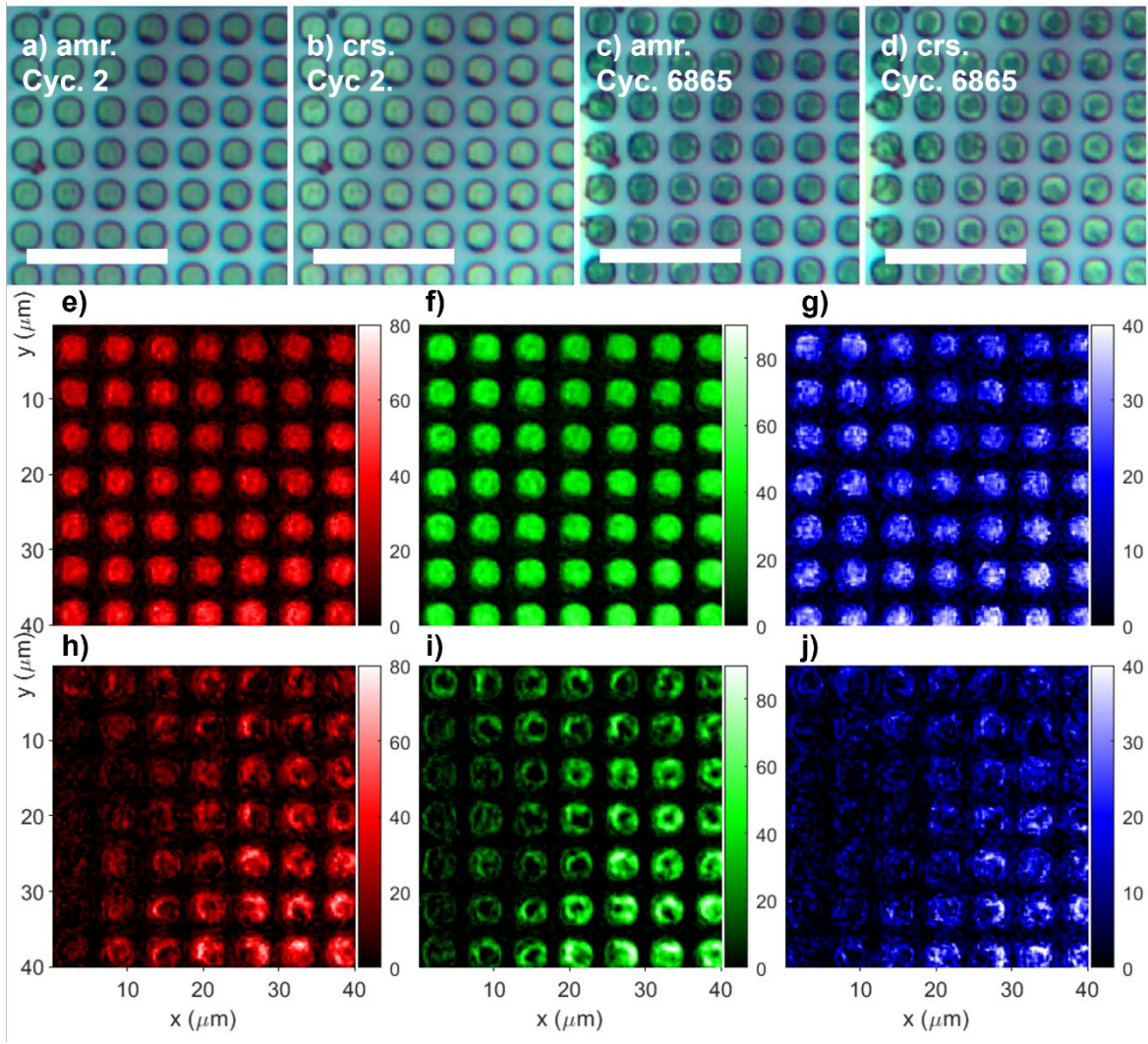

**Figure 4** (a-d) Optical micrographs of a GSST-integrated heater in amorphous (a and c) and crystalline states (b and d). (a & b) show uniform switching from initial cycles of the sample and (c & d) imply non-uniform switching due to partial failure after 6,865 switching cycles of the device. The scale bar is 20 μm for a-d. (e-j) Pixel intensity contrasts across the R, G and B color channels between (e-g) images a and b, and (h-j) images c and d, respectively.

As an example, **Figure 4** presents data collected from two samples where the PCM films are patterned into cylinder arrays. Figure 4a-b shows spatially uniform switching of PCM evidenced by the contrast maps in Figure 4e-g. In comparison, Figure 4c-d and Figure 4h-j show data collected on a sample which has undergone 6,865 switching cycles and has partially failed, as only selected PCM unit cells (at the bottom right corners of the images) continued to switch. The device can then be sectioned with focused ion beam and analyzed with TEM. Differences in microstructures and composition gradients between the "active" and "dead" unit cells provide valuable insights elucidating the mechanisms leading to PCM failure (the details of which are summarized in a forthcoming publication). **Figure 5** shows the measured DMA on a sample designed with enhanced reliability, indicating consistent switching over more than 10,000 cycles.

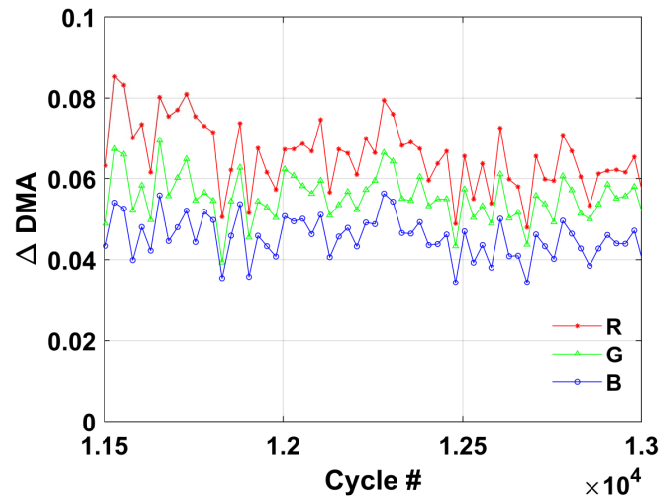

**Figure 5** DMA parameter evolution over cycles in a GSST device with 180 nm thick GSST and 740 nm thick SiN$_x$ encapsulation.

## 4. Conclusion

In this article, we present a multi-functional platform capable of characterizing the structural, optical, kinetic, thermal and cycling behavior of optical PCMs. The platform is based on SOI microheater arrays and an electrothermal switching scheme representative of the PCM deployment environment when integrated with industry-standard silicon photonic devices. We demonstrate that the platform enables *in situ*, quantitative characterizations of the phase composition, optical constants, temperature profiles, TTT diagram, cycling endurance, and material uniformity of PCM thin films. It can also be coupled with combinatorial deposition to facilitate high-throughput screening of new PCM compositions, or combined with *ex situ* characterization techniques such as electron microscopy to elucidate the microstructural evolution and failure mechanisms of PCMs. The microheater array devices can be fabricated leveraging commercially available photonic foundry services, and we open-sourced hardware and software designs of the characterization platform, allowing the broader PCM community to make use of its unique capabilities. It is our hope that the work will expand and expedite discovery and qualification of new optical PCMs for diverse photonic applications.

## 5. Acknowledgements

Funding support is provided by NSF under award number 2132929 and by NASA under contract 80LARC17C0004. B.Mills also acknowledges support provided by the Draper Fellowship. C. Ríos acknowledges support from the NSF under Grant ECCS-2210168.

## 7. Table of Contents

Table of Contents entry: Chalcogenide phase change materials show great promise for tunable nanophotonic devices. In this paper, an open-source electrical switching platform enabling in-situ, high-throughput, and multi-modal testing of phase change materials is presented. The platform will accelerate discovery and characterization of new phase change materials and facilitate their integration with on-chip photonic devices.

Cosmin-Constantin Popescu, Khoi Phong Dao, Luigi Ranno, Brian Mills[†], Louis Martin, Yifei Zhang, David Bono. Brian Neltner, Tian Gu, Juejun Hu*, Kiumars Aryana, William M. Humphreys, Hyun Jung Kim*, Steven Vitale, Paul Miller, Christopher Roberts, Sarah Geiger, Dennis Callahan, Michael Moebius, Myungkoo Kang, Kathleen Richardson, Carlos Ríos

**Title** An Open-Source Multi-functional Testing Platform for Optical Phase Change Materials

Table of Contents figure;

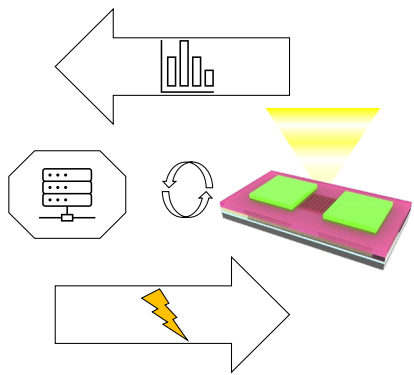

# Supporting Information

**Title** An Open-Source Multi-functional Testing Platform for Optical Phase Change Materials

*Cosmin-Constantin Popescu, Khoi Phuong Dao, Luigi Ranno, Brian Mills[†], Louis Martin, Yifei Zhang, David Bono. Brian Neltner, Tian Gu, Juejun Hu\**

Department of Materials Science & Engineering, Massachusetts Institute of Technology Cambridge, MA 02139, USA

E-mail: hujuejun@mit.edu

[†] Draper Scholar, The Charles Stark Draper Laboratory, 555 Technology Square, Cambridge, MA 02139, USA

*Kiumars Aryana, William M. Humphreys, Hyun Jung Kim\**

NASA Langley Research Center, Hampton, VA 23681, USA

E-mail: hyunjung.kim@nasa.gov

*Steven Vitale, Paul Miller, Christopher Roberts*

Advanced Materials and Microsystems Group, MIT Lincoln Laboratory Lexington, MA 02421, USA

*Sarah Geiger, Dennis Callahan, Michael Moebius*

The Charles Stark Draper Laboratory, Inc., Cambridge, MA 02139, USA

*Myungkoo Kang, Kathleen Richardson*

CREOL, The College of Optics & Photonics University of Central Florida Orlando, FL 32816, USA

*Carlos A. Ríos Ocampo*

Department of Materials Science & Engineering University of Maryland College Park, MD 20724, USA

The conductivity of doped silicon was estimated starting with a device resistance of 44 Ω at room temperature and assuming that the resistance corresponded to the doped heater region solely. The voltage and current reported by the voltage source were recorded and the corresponding resistance computed. For the conductivity, a 150 μm square doped Si heater with 116 nm SOI thickness was used.

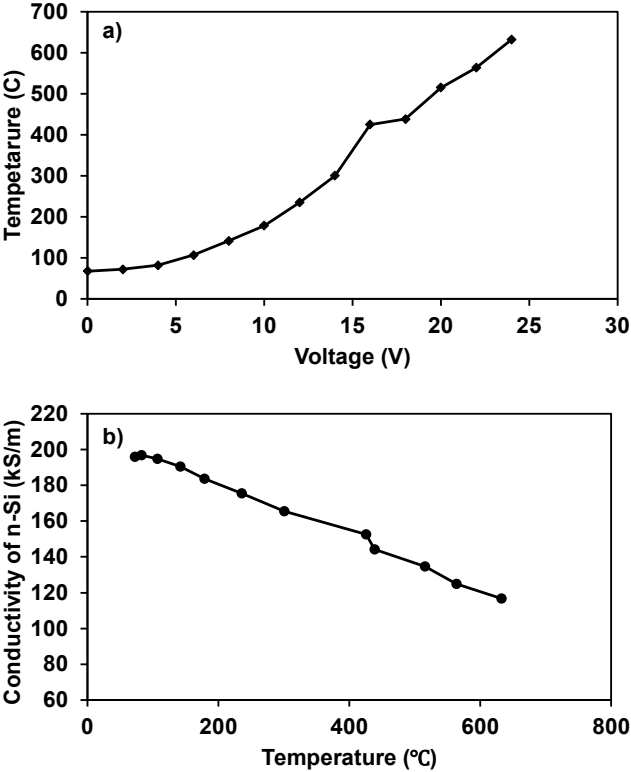

Figure S1 a) Temperature estimated from Raman spectroscopy measurements vs. voltage and b) calculated doped silicon conductivity vs. temperature